# Stochastic model of business process decomposition

*Grigory Tsiperman*

*Assistant professor*
*National University of Science and Technology "MISIS"*

gntsip@gmail.com

**Abstract**

Decomposition is the basis of works dedicated to business process modelling at the stage of information and management systems' analysis and design. The article shows that the business process decomposition can be represented as a Galton–Watson branching stochastic process. This representation allows estimating the decomposition tree depth and the total amount of its elements, as well as explaining the empirical requirement for the business function decomposition (not more than 7 elements). The problem is deemed relevant as the obtained results allow objectively estimating the labor input in business process modelling.

## 1. Introduction

At the initial development stage, it is difficult to predict the project complexity, which is determined by the number and structure of ties between the subject field facts, which cause respective requirements to the information system functionality. However, for the information system development practice, this issue is particularly relevant due to the necessity to ensure the fair project value. An important part of the analysis and design is the study and modelling of the business processes that are to be automated.

Business process modelling is the basis of the development technology for requirements to information systems, building and analyzing management systems. The complexity of the project can be assessed by the complexity of its model. In a number of works [1, 2, 3] it was suggested to use the concept of entropy as a measure of the complexity of business processes. This approach is applicable to the evaluation of small fragments of business processes, generating recommendations for their improvement. However, it seemed impossible to find literature that appeared to be natural metrics related to the number of elements of the business process model. This paper is an attempt to introduce such metrics.

The main business process modelling method is, as a rule, the decomposition method, that is the downward business process description, from general functions to more detailed ones. It is reasonably assumed that the number of the decomposition tree levels, its depth, is the greater, the more complex the business process is. The number of decomposition elements (business functions) seems immensely large. Of course, it is possible to decrease the decomposition depth by increasing the average number of decomposition elements for each business function, but this will result in the increasing complexity of the model. Note that Douglas Ross set a rule in the requirements for the business process modelling in SADT [4]: a business function decomposition should have at most 7 elements. This requirement ensured obtaining intelligible business process models.

We consider this thesis in more detail. Let us assume that the general business process decomposition tree allowable for the design and analysis cannot have more than $T$ decomposition elements (we treat this parameter as a resource limit). We introduce two more parameters: $\lambda$ is the average number of the business function decomposition elements, and $K$ is the decomposition tree depth. In this case

$$T = \frac{\lambda^{K+1}-1}{\lambda-1}.$$

Hence it follows that:

$$K = \left\lceil \frac{\ln(1+(\lambda-1)T)}{\ln(\lambda)} \right\rceil - 1.$$

As one can see from the graph of the function calculated for $T = 1000$ (Figure 1) the decomposition depth decreases as $\lambda$ increases.

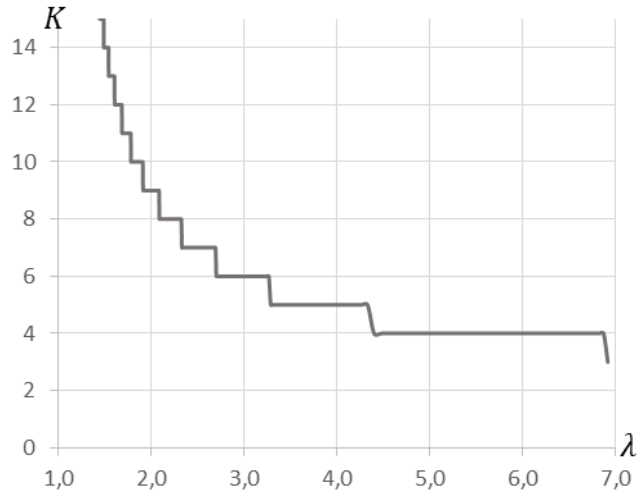

Figure 1. The decomposition depth ($K$) dependence on the average number of the decomposition elements ($\lambda$)

However, can we assume a priori what number of elements will a model have? The fact is that $T$ is a random variable by its nature.

And how the decomposition depth depends on the expectation of the business function decomposition elements' number?

My personal experience as well as the experience of my colleagues show that whatever the efforts, the decomposition tree depth for majority business processes does not exceed 4 or 5. Why is that?

On top of that: why does the magic number 7 suggested by Douglas Ross work?

We will try to find answers to these questions.

## 2. Business process decomposition

Let us consider an example of decomposition of one of the most trivial business processes: the process of making fried eggs. In order to describe the action sequence, we need to build a certain theoretic structure of objects, which jointly answers the questions "What to do?" and "How to do?". In other words, we need to construct a model for the egg frying process.

We start with the object that designates the "Make fried eggs" process. The next level objects detail actions required for implementing this process. We call these objects 'business functions'.

Generally, the business process itself is of the same nature as a business function. In another, more general, business process (e.g. "Make breakfast"), this process would be also called

'business function' (Figure 2). We separate the business process from business functions in order to mark the starting point of the decomposition.

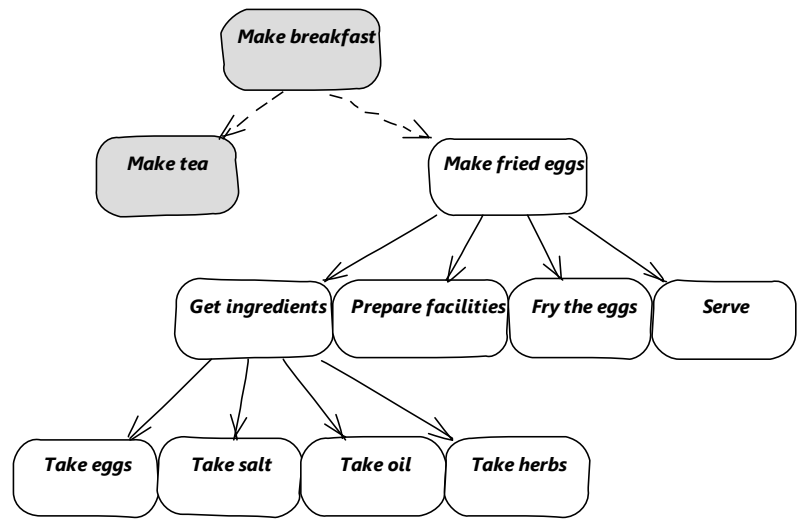

**Figure 2. The decomposition tree**

Each of the represented business functions can be as well decomposed. For example, business function "Get ingredients" can be represented as a subset of business functions, such as: 'take eggs', 'take salt', 'take oil (for frying)', 'take herbs (for serving)', etc.

Normally, a business process decomposition model is a tree, the root being the name of the business process, and its vertices being business functions.

A business process decomposition procedure is described by three main properties: variability, intelligibility, and finiteness. Let us consider them in more detail.

## 2.1. Decomposition variability

Variability is defined as the possibility of representing the decomposition of any business function with various business function sets. In other words, variability specifies the random nature of selection of a subset of business functions decomposing the original business function.

For example, it is possible for the example under consideration to combine at the first level of our decomposition (Figure 3) the business functions of preparing ingredients and tools for making fried eggs, by naming the new business function 'Prepare for making fried eggs'. At the next stage we can decompose this function into the stages of selecting ingredients and tools. Selection of business functions depends on various factors: the purpose of the business process description, the description intelligibility by means of reducing the number of business functions at each level of decomposition, and finally, the developer's preferences.

An important factor of variability is independence of the number of the decomposition elements of any business function from the previous decomposition of the business process. The decision on the decomposition representation depends only on the selected detail level for the business function description and on combination of the resulting elements of the decomposition.

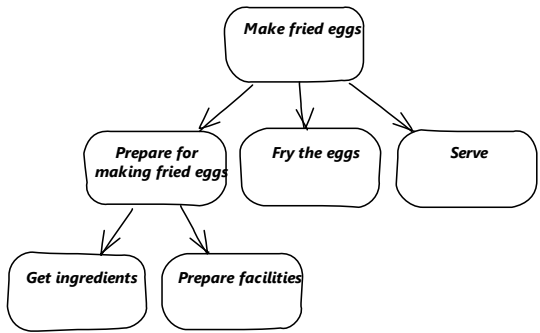

**Figure 3. Decomposition variability**

## 2.2. Decomposition intelligibility

Discussing the business process decomposition, Douglas Ross, the SADT author, considered the requirement to ensure the model representation intelligibility to be one of the basic requirements [4]. He formulated the empirical rule, according to which any business function decomposition cannot include more than 7 elements.

Decomposition is not the only method for business process representation. In the limiting case, it can be described as a sequence of operations resulting in obtaining the required result. However, such description is hard to understand and analyze in case of real processes that involve hundreds of operations.

Decomposition as a method of business process model representation creates a classification scheme combining operations in accordance with the system of intermediate results of the business process. As noted by Ross, a well-elaborated functional decomposition is mainly based on proper decomposition of the system objects.

Thus, the measure of a model intelligibility is the number of elements of the business function decomposition. Taking into account the decomposition variability property, one can always ensure the boundedness of the quantity.

For example, let us assume that a decomposition of a business function $A$ includes too many elements (in SADT terms):

$A \Longrightarrow (A_1, A_2, \ldots, A_n), 7 \leq n \leq 14$

We can introduce an additional intermediate grouping by decomposing $A$ into a fewer number of business functions, which allow for further decomposition, that is:

$A \Longrightarrow (A', A'', A''')$,
$A' \Longrightarrow (A_1, \ldots, A_k), k \leq 7$,
$A'' \Longrightarrow (A_{k+1}, \ldots, A_m), m - k \leq 7$,
$A''' \Longrightarrow (A_{m+1}, \ldots, A_n), n - m \leq 7$.

## 2.3. Decomposition finiteness

One of the important problems to be solved by a business process decomposition is to specify the decomposition horizon, i.e. a finite level, at which the decomposition of business functions should be stopped. The maximum number of the finite decomposition level is called the business process decomposition depth. In other words, the decomposition depth is the maximum distance from the root to the leaves of the decomposition tree.

When a model is being built, the decomposition depth, its horizon, is determined by either the point of view, based on which the description is provided, or the subject area, within which the business process is described. Let us clarify this thesis:

The description of a business process from the viewpoint of an enterprise executive should be ended before the description of the production of individual components. Business functions that would describe this production in detail are important for a shop foreman, but are out of scope of immediate control of the enterprise executive.

Theoretically, the decomposition depth may be very large: any action can be decomposed physiologically, then chemically and physically. However, it is absolutely clear that in this case we go out of the description of the business process itself and shift to another subject field that is out of the scope of our interest.

The more detailed is the desirable level of description of a business function, the smaller the actions are, to which it needs to be decomposed. For example, in our case the function 'Prepare for cooking fried eggs' can be represented in detail by specifying all required ingredients and tools. Thus, the more detailed is the description, the more elements of the business function decomposition are attracted.

One more example. In developing information systems, the business process decomposition should end at the level of determining the production operations that are to be automated. From this point, each operation should be described in terms of those information services that are to be applied in it. In this case, we proceed to another subject area, for which its own hierarchy of objects occurs, describing the developed information system.

Of course, the overall decomposition picture in this case is represented by a single tree, but the nature of the tree's elements can vary.

## 2.4. Statement of the problem

To summarize the description of the business process decomposition properties, we can say that the formulation of the problem for the decomposition process model development assumes the following:

1) The description of **a** decomposition model should be made in terms of a stochastic Markov process.

2) The decomposition process model should predict the following probabilistic characteristics:
   - the decomposition depth;
   - the total number of the business process model decomposition elements.

## 2.5. The probabilistic decomposition model

With account of the variability property, the decomposition process can be treated as a stochastic Markov process. Let us consider the formal definition of the process [5]:

Let $\mathcal{B} = \{f_i\}$, where $i \in \mathbb{N}$ be the set of all business functions $f_i$ related to a particular subject area we are interested and let $B \subset \mathcal{B}$ be a subset of $\mathcal{B}$. We specify the decomposition relation:

$$D_i = D(f_i): f_i \to B_i = (f_{i1}, f_{i2}, \dots, f_{in}), f_i \notin B_i \quad (1)$$
$$\forall i, j \in \mathbb{N}: B_i \cap B_j = \emptyset.$$

We say that a business function $f_g$ belongs to the decomposition horizon if

$$D(f_g) = \emptyset. \tag{2}$$

Using the definitions above we can represent a decomposition process as follows. Initially, we have one element designated by the name of the business process under decomposition:

$$Z(0) = 1. \tag{3}$$

As a result of this element decomposition, a random number $X$ of the decomposition elements of the first generation (level) is generated due to the variability:

$$X_1^{(0)} = |D(f_0)| = |B_0 = (f_{01}, f_{02}, \ldots, f_{0Z(1)})|. \tag{4}$$

In (4), $X_1^{(0)}$ is the number of elements of the original element decomposition. The subsequent decompositions are constructed by independent decompositions of each element of the previous level. In a formal way, the structure of the decomposition can be described as follows:

$$\begin{aligned}
D_0 &= D(0) = (f_0), \\
D_1 &= D(f_0) = (f_1, f_2, \ldots, f_n), \\
D_{21} = D(f_1) &= (f_{11}, \ldots, f_{1m}); \ldots; D_{2n} = D(f_n) = (f_{n1}, \ldots, f_{np}), \\
&\ldots\ldots\ldots \\
D_{\underbrace{ki\ldots j}_{k}} &= D\left(f_{\underbrace{i\ldots j}_{k-1}}\right) = \left(f_{\underbrace{i\ldots j1}_{k}}, \ldots, f_{\underbrace{i\ldots jp}_{k}}\right).
\end{aligned} \tag{5}$$

The number of elements of the decomposition of an element $f_{i\ldots j}$ at the $k^{th}$ level is determined as:

$$X_{i\ldots j}^{(k)} = |D_{ki\ldots j}|. \tag{6}$$

Then the number of elements in the $k^{th}$ generation of the decomposition is determined by the expression

$$Z(k) = \sum_j X_{i\ldots j}^{(k)}. \tag{7}$$

Observe that, if $f_{i\ldots j}$ belongs to the decomposition horizon, then:

$$X_{i\ldots j}^{(k)} = 0. \tag{8}$$

### 2.5.1. Probability space of the decomposition process

We call a tuple

$$\omega = (f_0, f_{1i}, f_{2ij}, \ldots, f_{nij\ldots p}) \tag{9}$$

an elementary implementation of the decomposition process.

The probability space $(\Omega, \mathcal{F}, P)$ on which the decomposition process is defined consists of the space of events, which is the set of all elementary implementations

$$\Omega = \{\omega\}. \tag{10}$$

The σ-algebra $\mathcal{F}$ is generated by the cylindrical subsets of $\Omega$, and the probability measure $P$ is described by the relations:

$$P\{X_{i...j}^{(k)} = n\} = p_n,$$
$$P\left\{\left(X_1^{(0)}, X_i^{(1)}, ..., X_{i...j}^{(k)}\right) = \left(n_{01}, n_{1i}..., n_{ki...j}\right)\right\} = p_{n_{01}} p_{n_{1i}} ... p_{n_{ki...j}}. \tag{11}$$

where $\sum p_n = 1, p_n \geq 0$.

### 2.5.2. Decomposition as a branching process

If we assume the independence of each decomposition from the previous decompositions, the sequence of random variables $Z(0), Z(1), ...$ constitutes a Markov chain. The decomposition process time scale is discrete, that is the sequential time moments correspond to the beginning and end of the business function decomposition. Therefore, we can assume that the life time of any decomposed vertex of the decomposition tree is equal to 1.

In this case, the business process decomposition can be treated as a Galton–Watson branching process.

We assume that our decomposition is described by a supercritical branching process with extinction probability $\alpha > 0$. The supercritical nature of the process means that the expected number of the decomposition elements of each business function exceeds 1, that is

$$\lambda = E[X] > 1. \tag{12}$$

### 2.5.3. The probability distribution for the number of decomposition elements

To find the probability distribution of a random variable $X$, which is the number of elements of a business function decomposition, we show that the flow of the decomposition elements' occurrence events can be interpreted as a simple stationary and ordinary flow having independent increments.

A decomposition can be assumed to take place within the time period [0,1] and consists of a small random number of elements. Taking into account the peculiarities of the decomposition process, we can affirm that the elements occur in this period individually at random time points. More precisely, the probability that an event occurs within an interval $\Delta t$ is equal to $\lambda \Delta t + o(\Delta t)$, where $\lambda$ is the flow intensity, and the probability of occurrence of more than one event within this interval is $o(\Delta t)$. This corresponds to the ordinarily flow of events.

The stationarity property means that the probability of occurrence of an event depends only on the time interval $\Delta t$ and does not depend on the time point $t$. This is true for the decomposition process, as the decomposition elements, which are also business functions, do not depend on each other, and, accordingly, each element can occur at a random time point within the life time of the decomposed element.

Finally, the decomposition elements' independence provides independence for the number of events occurring at non-overlapping life time intervals of the decomposition process.

As it is shown in [6], the probability of occurring $n$ events within the time interval $\Delta t$ is determined by the Poisson distribution, i.e.

$$P\{X(\Delta t) = n\} = ((\lambda \Delta t)^n / n!) \exp(-\lambda \Delta t). \tag{13}$$

Taking into account that we are interested in the probability distribution for the number of decomposition elements of the business function, when $\Delta t = 1$, (13) will have the following form:

$$P\{X = n\} = (\lambda^n / n!) \exp(-\lambda), \tag{14}$$

where $\lambda$ is the expectation of the number of the business function decomposition elements, corresponding to condition (12).

## 3. The quantitative estimation of decomposition

### 3.1. The decomposition process model

Let us consider the following decomposition process model (see Figure 4).

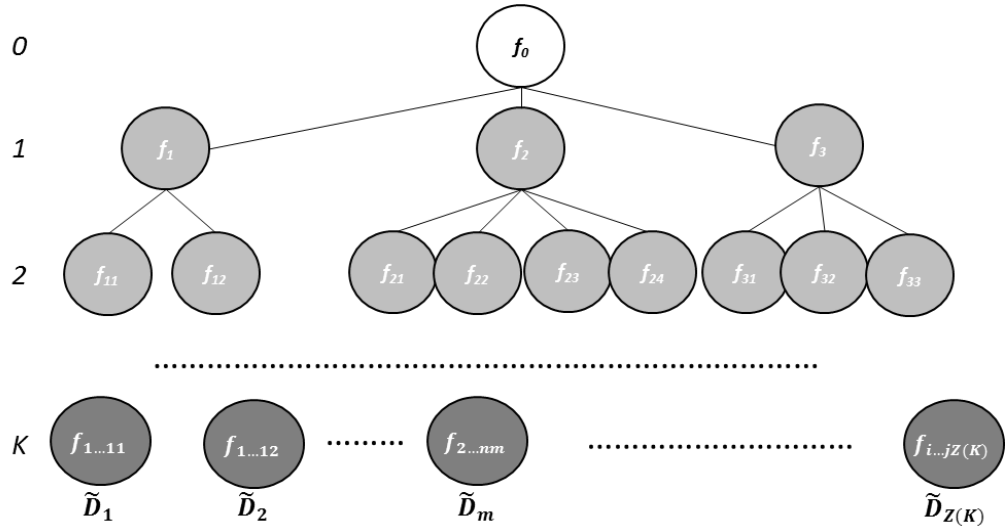

**Figure 4. The structure of a decomposition process**

We will consider the horizon of the decomposition as the time moment preceding the extinction moment of a branching process.

Assume that the horizon of a decomposition process is located at a certain level $K$. The number of decomposition elements at this level is a random variable $Z(K)$, the expectation of which will be denoted as $E[Z(K)]$. It is clear that the lesser the number of elements at the level is, the higher the probability of the event that all the elements will not be decomposed simultaneously, that is, the number of the decomposition elements at level $K$ should not be too large to make the probability of degeneration negligibly low.

**We** consider an element of this level as the vertex of **a** degenerating subprocess of decomposition $\widetilde{D}_i$. The expected number of such subprocesses is equal to $E[Z(K)]$. We use $\Delta n$ to designate the number of the decomposition elements generated within the subprocess $\widetilde{D}_i$. Here and in the sequel, all variables related to subprocesses $\widetilde{D}_i$, will be supplied by tilde.

To complete the decomposition process at level $K$ we require that the expected number of the decomposition elements meets the condition

$$\Delta N = E[Z(K)] \Delta n < 1. \tag{15}$$

In this case, $K$ determines the maximum possible decomposition horizon. Decomposition of a business process may be stopped at earlier levels. It depends on various factors, including the task statement, the degree of the process understanding by a designer, and even on the limitation of the design time.

To provide the validity of condition (15), it is natural to assume that each of subprocesses $\widetilde{D}_i$ dies out already at the second generation meaning that

$$\Delta n = E[\widetilde{Z}(1); \widetilde{Z}(2) = 0] = \sum_{i=1}^{\infty} i P\{\widetilde{Z}(1) = i; \widetilde{Z}(2) = 0\}. \tag{16}$$

## 3.2. The expected decomposition depth

Since our subsequent arguments are based on the theory of branching processes, we recall some basic notions of the theory.

A branching process is specified by a probability generating function

$$f_n(s) = \sum_{i=0}^{\infty} P\{Z(n) = i\} s^i, \quad 0 \leq s \leq 1; n \geq 0. \tag{17}$$

Here $f_n(s)$ is the probability generating function for the number of particles of the $n^{th}$ generation of the branching process, and $P\{Z(n) = i\}$ is the probability that this generation has precisely $i$ elements [5]. Thus, the probability of extinction the process to the $n^{th}$ generation is

$$f_n(0) = P\{Z(n) = 0\}. \tag{18}$$

The probability generating functions for different generations of the branching process are iteratively connected

$$f_0(s) = s, \quad f_{n+1}(s) = f(f_n(s)), \tag{19}$$

where $f(s) = f_1(s)$.

In the sequel we need the probability generating function $f(s)$ for Poisson distribution [7] with parameter $\lambda$:

$$f(s) = f_1(s) = \sum_{i=0}^{\infty} P\{Z(1) = i\} s^i = \sum_{i=0}^{\infty} \frac{(\lambda s)^i}{i!} e^{-\lambda} = e^{\lambda(s-1)}. \tag{20}$$

In this case the probability generating function for the number of particles of the $n^{th}$ generation looks as follows:

$$f_n(s) = e^{-\lambda} \underbrace{exp\left(\lambda e^{-\lambda} exp\left(\lambda e^{-\lambda} exp\left(\ldots exp\left(\lambda e^{-\lambda} exp(\lambda s)\right)\right)\right)\right)}_{n \; times}. \tag{21}$$

### 3.2.1. The branching process conditioned on extinction

Let us consider a general problem of finding the expected number of particles in the $n^{th}$ generation of a branching process given its extinction at the $(n+1)^{th}$ level. It is necessary to evaluate the quantity

$$E[\tilde{Z}(n); \tilde{Z}(n+1) = 0] = \sum_{i=1}^{\infty} iP\{\tilde{Z}(n) = i; \tilde{Z}(n+1) = 0\} \qquad (22)$$

for any $n$ or, taking into account dependence of $\tilde{Z}(n)$ and $\tilde{Z}(n+1)$,

$$E[\tilde{Z}(n); \tilde{Z}(n+1) = 0] =$$
$$= \sum_{i=1}^{\infty} iP\{\tilde{Z}(n) = i\}P\{\tilde{Z}(n+1) = 0 | \tilde{Z}(n) = i\}. \qquad (23)$$

Consider the subprocesses initiated by particles of the $n^{th}$ generation. The generation number, in which each of these subprocesses dies out, is equal to 1. Since all the $\tilde{Z}(n)$ subprocesses are independent, the probability of simultaneous extinction of all these subprocesses is equal to the product of the extinction probabilities of each of them. Thus, the conditional probability at the right hand side of (22), can be represented in the following form:

$$P\{\tilde{Z}(n+1) = 0 | \tilde{Z}(n) = i\} = \left(P\{\tilde{Z}(1) = 0 | \tilde{Z}(0) = 1\}\right)^i = f_1(0)^i. \qquad (24)$$

We have used in (24) the probability of extinction of a branching process given in (18). Substituting (24) in (23), we find:

$$E[\tilde{Z}(n); \tilde{Z}(n+1) = 0] = \sum_{i=1}^{\infty} iP\{\tilde{Z}(n) = i\}f_1(0)^i = f_1(0)f_n'\big(f_1(0)\big). \qquad (25)$$

### 3.2.2. The maximum decomposition horizon

Using (25), we now determined $\Delta n$ defined by (16). Clearly,

$$\Delta n = E[\tilde{Z}(1); \tilde{Z}(2) = 0] = f_1(0)f_1'\big(f_1(0)\big). \qquad (26)$$

Recalling expression (21) for the probability generating function of Poisson distribution, we transform (26) to the following form:

$$\Delta n = \exp(-\lambda)\left(\lambda\exp(\lambda(\exp(-\lambda) - 1))\right) = \lambda\exp(\lambda(\exp(-\lambda) - 2)). \qquad (27)$$

On account of (27), condition (14), giving the signal to stop the decomposition process, transforms to the following form:

$$\Delta N = \lambda^{K+1}\exp(\lambda(\exp(-\lambda) - 2)) < 1. \tag{28}$$

Estimate (28) takes into account the fact that the expected number of the decomposition elements at level $K$ is equal to $\lambda^K$.

Finding the logarithm of (28) and solving the obtained inequality with respect to $K$, we arrive to the following estimate for the maximum nonempty generation of the decomposition process:

$$K < g(\lambda) = \lambda\,(2 - \exp(-\lambda))/\ln\lambda - 1, \tag{29}$$

where $g(\lambda)$ is the function determining the maximum horizon.

Thus, the maximum decomposition horizon is specified as the maximal positive integer not exceeding $g(\lambda)$:

$$K = \lfloor \lambda\,(2 - \exp(-\lambda))/\ln\lambda - 1 \rfloor. \tag{30}$$

### 3.2.3. The expected decomposition horizon

As we have already noted, a decomposition can be stopped at any level before the maximum one. Let us consider the decomposition horizon as a random variable $G(K)$ that may take values from 1 to K. The aim is to estimate the distribution law $P\{G(K)\}$.

Let us find a linear approximation for $P\{G(K)\}$ using the following considerations. We assume that

$$P\{G = 0\} = 0 \text{ and } P\{G = K + 1\} = 0. \tag{31}$$

These assumptions are quite natural. Indeed, it has no sense to stop the decomposition process at level zero. Similarly it is also natural to assume that the probability of attaining a horizon exceeding the maximum is negligibly small.

We take a horizon close to the middle of interval $[0, K + 1]$ as an expected horizon. Experience shows that horizon $G = 3$ is the most probable for $\lambda < 7$. Therefore, we assume that the mode of the distribution corresponds to the horizon value:

$$K_m = \lceil (K + 1)/2 \rceil = \begin{cases} (K + 2)/2 & \text{for even } K; \\ (K + 1)/2 & \text{for odd } K. \end{cases} \tag{32}$$

Basing on conditions (31) and (32), we can easily find a linear approximation for the distribution law of the decomposition horizon:

if $K$ is even, then

$$P\{G = n\} = \begin{cases} 4n/((K+1)(K+2)), & 0 \le n \le K_m, \\ 4(K+1-n)/(K(K+1)), & K_m + 1 \le n \le K + 1, \end{cases}$$

if $K$ is odd, then

$$P\{G = n\} = \begin{cases} 4n/(K+1)^2, & 0 \le n \le K_m; \\ 4(K+1-n)/(K+1)^2, & K_m + 1 \le n \le K + 1. \end{cases} \tag{33}$$

Using these expressions we have calculated the expected decomposition horizon for $2 \le \lambda \le 12$ (Figure 5):

$$\bar{K} = \left\lceil \sum_{n=1}^{K+1} nP\{G = n\} \right\rceil = \begin{cases} 3 \text{ if } 2 \leq \lambda \leq 6.7; \\ 4 \text{ if } 6.7 < \lambda \leq 10.7; \\ 5 \text{ if } 10.7 < \lambda \leq 12. \end{cases} \qquad (34)$$

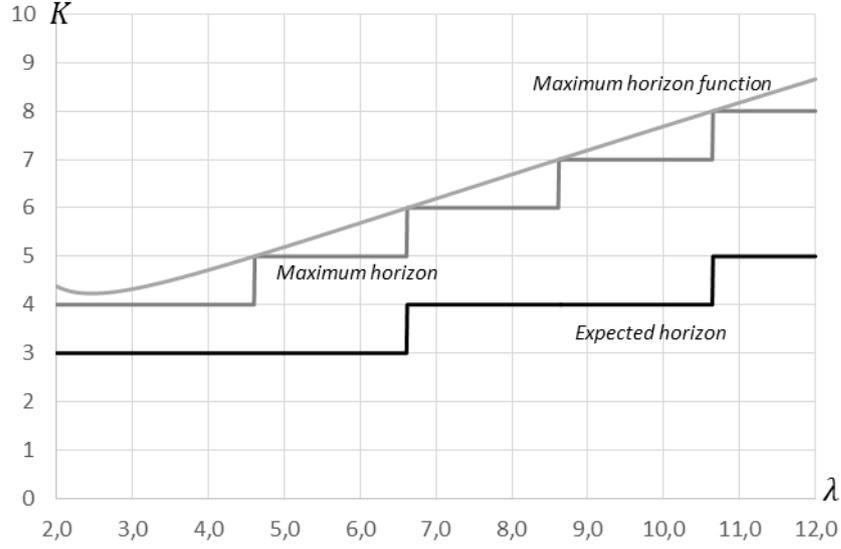

**Figure 5. Decomposition horizon**

## 3.3. Estimation for the total number of the decomposition elements

To estimate the labor input, it is necessary to evaluate the total number of the decomposition process elements. Let us consider two scenarios of such estimation.

The first scenario concerns the case of full a priori indefiniteness of the decomposition result, when it is not possible to predict in advance the number of the model's levels. In other words, the scenario predicts the general number of elements under a random decomposition depth.

The second scenario is based on a more realistic estimate of the decomposition depth, i.e. when it is clear in advance, which horizon is expected. This situation occurs in the vast majority of cases. This scenario predicts the total number of the decomposition elements at an expected horizon, determined by (34).

### 3.3.1. The total number of elements at a random horizon

We assume that the decomposition depth is a random variable $G$, which takes values in the interval $(0; K + 1)$, where $K$ is determined by expression (30). A linear approximation to the probability distribution function $P\{G = n\}$ is specifies by (33).

Let $T(n)$ be the total number of the decomposition elements given the horizon $n$ and $T(G)$ be the total number of the decomposition elements with a random horizon.

As it was shown in section 3.2, the decomposition depth weakly depends on $\lambda$. Thus, we can assume that the total number of elements of decomposition $T(G)$ and its horizon are independent random variables. In other words,

$$P\{T(n) = j | G = n\} = P\{T(n) = j\} P\{G = n\}. \qquad (35)$$

In this case, the probability that as the total number of elements of the decomposition will be equal to $j$ is calculated by the formula

$$P\{T(G) = j\} = \sum_{n=1}^{K} P\{G = n\}P\{T(n) = j\}. \tag{36}$$

Therefore, the generating function for the total number of elements of decomposition $T(G)$ has the following form:

$$C_K(s) = \sum_{n=1}^{K} \sum_{j=1}^{\infty} P\{G = n\}P\{T(n) = j\} s^j = \sum_{n=1}^{K} P\{G = n\}R_n(s), \tag{37}$$

where $R_n(s)$ is the probability generating function for the total number of elements $T(n)$ in a process with the decomposition depth $n$. A particular form of $R_n(s)$ is not important for our consideration (a recursive formula for this function is provided in [8]). Our task is to determine the mean and variance of the total number of particles $T(G)$ in the decomposition process. They are calculated by the following formulas:

$$C_K'(s)_{|s=1} = E[T(G)] = \sum_{n=1}^{K} P\{G = n\}R_n'(s)_{|s=1} =$$
$$= \sum_{n=1}^{K} P\{G = n\}E[T(n)], \tag{38}$$

$$C_K''(s)_{|s=1} = E[T^2(G)] - E[T(G)] =$$
$$= \sum_{n=1}^{K} P\{G = n\}R_n''(s)_{|s=1} = \sum_{n=1}^{K} P\{G = n\}(E[T^2(n)] - E[T(n)]). \tag{39}$$

In (38), the expectation of the total number of elements contained in a decomposition with depth $n$ is calculated as the sum of expectations for the number of elements in each generation of the process:

$$E[T(n)] = \sum_{i=0}^{n} E[Z(i)] = (\lambda^{n+1} - 1)/(\lambda - 1). \tag{40}$$

Then, using (38), we can easily find the following expression for $E[T(G)]$:

$$E[T(G)] = \sum_{n=0}^{K+1} P\{G = n\}(\lambda^{n+1} - 1)/(\lambda - 1). \tag{41}$$

Taking into account the known formula for the variance of a random variable $X$

$$D[X] = E[X^2] - E^2[X], \tag{42}$$

we can rewrite (39) as follows:

$$D[T(G)] + E^2[T(G)] - E[T(G)] =$$
$$= \sum_{n=1}^{K} P\{G = n\}(D[T(n)] + E^2[T(n)] - E[T(n)]). \tag{43}$$

Hence, on account of (38), we find:

$$D[T(G)] = \sum_{n=1}^{K} P\{G = n\}(D[T(n)] + E^2[T(n)]) - E^2[T(G)]. \qquad (44)$$

Representation (44) allows us to calculate $D[T(G)]$ as all the components required for the calculations are known: $E[T(n)]$ is determined by expression (47), $E[T(G)]$ is determined by (41), the probability distribution for the decomposition horizon is given by (33). The variance $D[T(n)]$ for the total number of elements when the expected horizon of a decomposition is given by the expression

$$D[T(n)] = \lambda^n \frac{\lambda^n - 1}{\lambda - 1} + \frac{2\lambda + 1}{\lambda - 1}\left(\frac{\lambda^{2n} - 1}{\lambda^2 - 1} - \frac{\lambda^n - 1}{\lambda - 1}\right). \qquad (45)$$

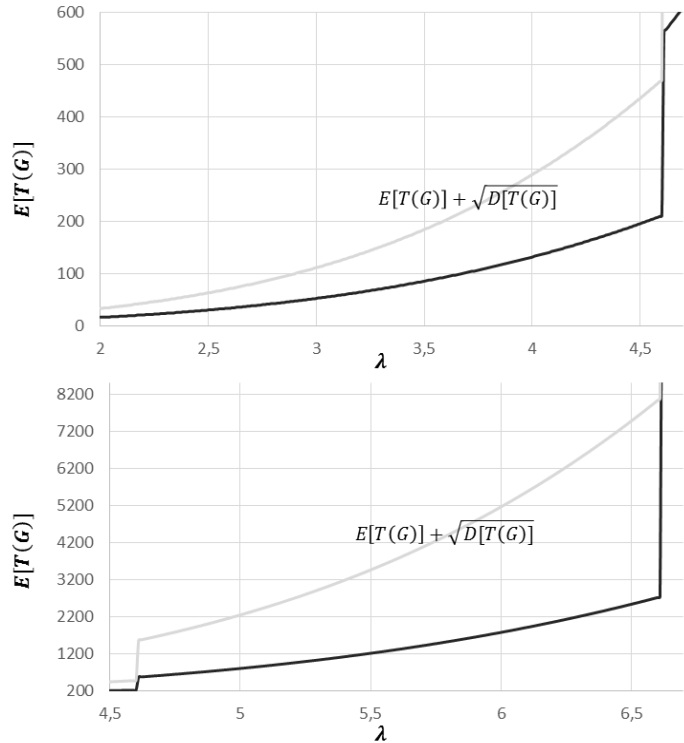

Figure 6. The total number of the decomposition elements at a random horizon

Figure 6 shows two jumps on the graphs at $\lambda = 4.628$ and $\lambda = 6.625$. These jumps correspond to the change in the expected decomposition horizon.

### 3.3.2. The total number of elements under expected horizon

The mean of the number of elements in the $n^{\text{th}}$ generation of a branching process (the business process decomposition level) is determined as

$$E[Z(n)] = \lambda^n, n = (0,1,2,\ldots) \qquad (46)$$

Let $T(n)$ be the total number of elements in a process with decomposition depth $n$. The estimate of the total number of the decomposition elements for horizon $\overline{K}$ can be find by the formula for the sum of elements of a geometric progression

$$E[T(\bar{K})] = \sum_{i=0}^{\bar{K}} E[Z(i)] = \frac{\lambda^{\bar{K}+1} - 1}{\lambda - 1}, \qquad (47)$$

where $\bar{K}$ is determined by expression (30).

Variance for the total number of elements in the case of the expected horizon of decomposition $\bar{K}$ is given in (45).

Graphs drawn by Figure 7 show dependence of the total number of the decomposition elements on the expected offspring number of decompositions within intervals $2 \leq \lambda \leq 6.7$ and $6{,}5 \leq \lambda \leq 10.7$.

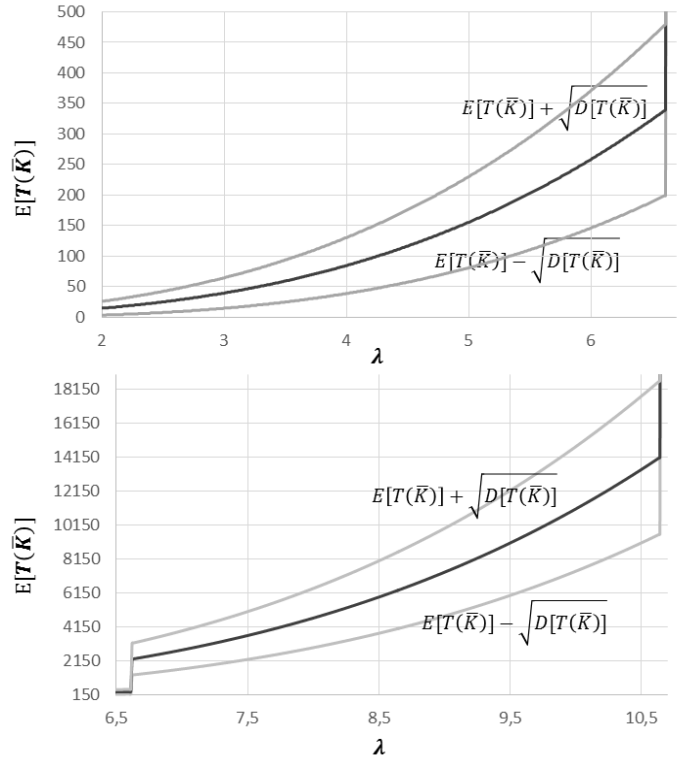

**Figure 7. The total number of the decomposition elements at an expected horizon**

The jumps on the graphs correspond to the changes in the expected decomposition horizon at $\lambda = 6.605$ and $\lambda = 10.655$.

## 4. Verification of the model based on experimental data

As the experimental data, we used the business process decompositions performed by the author in the following five information system projects:

1) a conceptual design of technically sophisticated facilities;
2) children distance learning;
3) an analytical system of an airline;
4) administration of a cloud service;
5) planning and monitoring production processes of shipbuilding companies.

For each project, a sampling of element decomposition distribution was formed (see Table 1), based on which the hypothesis on the Poisson distribution of the number of the decomposition elements was verified using the methods of statistical processing, as well as the general mean for a single decomposition.

| Project number | 1 | 2 | 3 | 4 | 5 |
|---|---|---|---|---|---|
| The number of elements in the decomposition | | | | | |
| 2 | 8 | 6 | 8 | 2 | 4 |
| 3 | 5 | 9 | 10 | 5 | 6 |
| 4 | 7 | 10 | 6 | 3 | 0 |
| 5 | 7 | 13 | 4 | 5 | 3 |
| 6 | 7 | 3 | 2 | 2 | 1 |
| 7 | 3 | 4 | 2 | 2 | |
| 8 | 3 | 3 | | 2 | |
| 9 | 4 | | | 1 | |
| 11 | 2 | | | 1 | |
| 12 | 1 | | | | |
| 13 | 1 | | | | |
| The total number of the decomposition elements | 264 | 214 | 117 | 118 | 48 |

**Table 1. The sampling of project-based decomposition magnitudes' distribution**

The verification of the Poisson decomposition distribution hypothesis brings positive results for all the projects.

Table 2 and Figure 8 provide a comparison of the results of statistical processing with model calculations.

| Project number | Experimental data | | | | | Model calculation data | | | |
|---|---|---|---|---|---|---|---|---|---|
| | Decomposition horizon | Mean for $\lambda$ | The lower margin for the confidence interval for $\lambda$ | The higher margin for the confidence interval for $\lambda$ | The total number of elements | Decomposition horizon | The lower margin of the confidence interval | Expected number of elements | The higher margin of the confidence interval |
| 1 | 4 | 5.41 | 4.58 | 6.24 | **264** | 2 - 4 | **61** | **194** | **412** |
| 2 | 4 | 4.46 | 3.97 | 4.95 | **214** | 2 - 4 | **38** | **114** | **214** |
| 3 | 3 | 3.55 | 3.01 | 4.09 | **117** | 2 - 4 | **15** | **61** | **138** |
| 4 | 3 | 5.13 | 4.12 | 6.14 | **118** | 2 - 4 | **43** | **167** | **394** |
| 5 | 3 | 3.2 | 2.41 | 3.99 | **48** | 2 - 4 | **7** | **47** | **130** |

Table 2. Comparison of the statistical processing results with model calculations

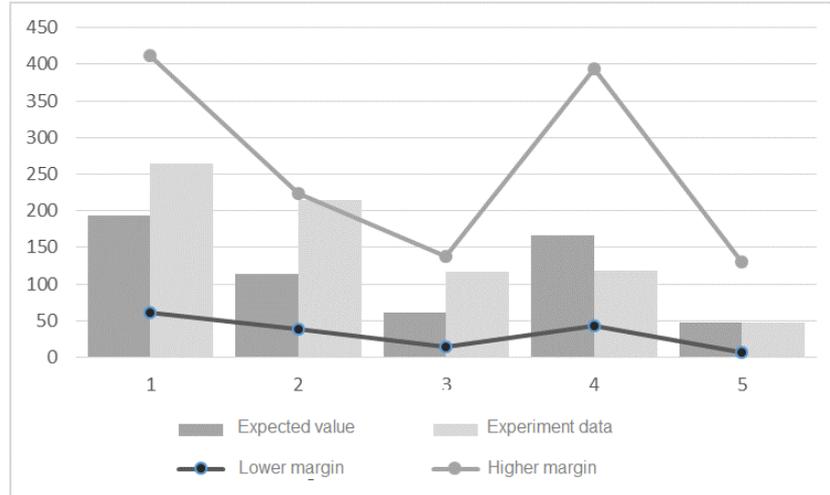

Figure 8. Representation of the correlation of experimental data with the calculation results

Thus, the experimental data for all the projects correspond to the model prediction by both the horizon and the total number of the decomposition elements.

## 5. Conclusion

As can be seen from the evaluation of the decomposition depth dependence provided in the introduction hereto (Figure 1) radically contradicts the real depth (Figure 5): the business process decomposition depth increases as the mean number of the business function decomposition elements grows. This phenomenon can be explained as follows:

The quality of a business process description is determined by the depth and level of detail, with which the analyst understands the modeled business process. And what is the measure of such understanding?

A business process decomposition is described by a supercritical Galton–Watson branching process, which is typically characterizes by positive probability of extinction $\alpha > 0$. Extinction probability is a measure of the business process understanding: the higher $\alpha$, the less detailed the business process description by the analyst is, and the faster the decomposition ends. Following [9], we introduce the complementary measure, the level of detail, $\gamma = 1 - \alpha$. The higher the level of detail is, the better the analyst describes the business process.

The extinction probability of a branching process is the minimal positive root of the equation

$$f(\alpha) = e^{\lambda(\alpha-1)} = \alpha.$$

Hence it follows that the mean of the decomposition elements number for the business function is

$$\lambda = -\frac{ln(\alpha)}{1-\alpha} = -\frac{ln(1-\gamma)}{\gamma}.$$

Dependence of $\lambda$ on the level of detail $\gamma$ is presented by Figure 9.

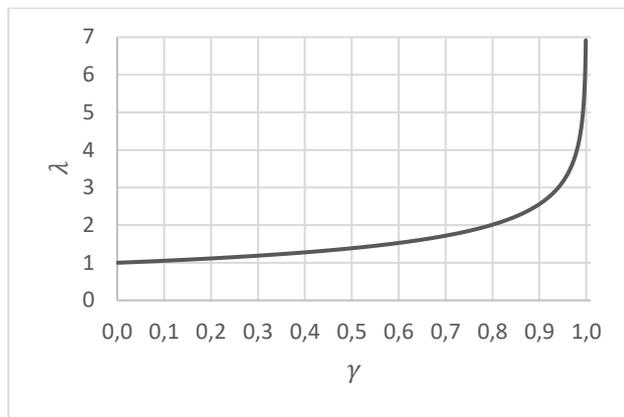

**Figure 9. The dependence of the expectation of the business function decomposition elements number on the level of detail**

Thus the grows of the decomposition depth along with the grows of the number of the decomposition elements is explained by a higher level of detail of the business process understanding by the analyst: the higher the level of detail is, the more complete the business process description is.

Figure 9 allows to understand one more phenomenon: the magic number 7, the maximum number of the business function decomposition elements. As one can see, the level of detail of the business process understanding is, at this point, close to 1, the maximum possible value and the graph looks here almost as a vertical line.

### Acknowledgments

The author expresses his sincere gratitude to Professor Vladimir Vatutin, a leading researcher of the Steklov Mathematical Institute of RAS, whose advices ensured the correct application of the methods of the branching processes theory and gave confidence in the validity of the obtained results.

# Стохастическая модель декомпозиции бизнес-процесса

*Циперман Григорий Наумович*

Доцент кафедры системной и программной инженерии МИСиС

**Аннотация**

Декомпозиция лежит в основе работ по моделированию бизнес-процессов на этапе анализа и проектирования информационных систем и систем управления. В статье показано, что декомпозицию бизнес-процесса можно представить как стохастический ветвящийся процесс Гальтона-Ватсона. Такое представление позволяет оценить глубину дерева декомпозиции и общее количество его элементов, а также объяснить эмпирическое требование декомпозиции бизнес-функции (не более 7 элементов). Актуальность задачи определяется тем, что полученные результаты дают возможность объективной оценки трудоёмкости работ по моделированию бизнес-процессов.

## *1. Введение*

Моделирование бизнес-процесса лежит в основе технологий проектирования требований к информационным системам, построения и анализа систем управления. Основным методом моделирования бизнес-процессов, как правило, является метод декомпозиции – описание бизнес-процесса «сверху вниз», от общих функций к более детальным.

На раннем этапе проектирования трудно предсказать сложность проекта, характеризуемую количеством и структурой связей между фактами предметной области, которые порождают соответствующие требования к функциональности информационной системы. Однако, для практики проектирования информационных систем этот вопрос имеет особую актуальность, вызванную необходимостью обеспечения объективной стоимости проекта. Важной частью работ по анализу и проектированию является изучение и моделирование автоматизируемых бизнес-процессов.

На уровне здравого смысла представляется, что число уровней дерева декомпозиции, его глубина, тем больше, чем сложнее бизнес-процесс, число элементов декомпозиции (бизнес-функций) кажется необозримо большим. Конечно можно уменьшить глубину декомпозиции, увеличив среднее количество элементов декомпозиции каждой бизнес функции, но это увеличит сложность модели. Кстати, Дуглас Росс в требованиях к моделированию бизнес процессов в SADT [1] установил правило: не более 7 элементов декомпозиции бизнес-функции. Это требование обеспечивало получение понятных моделей бизнес-процесса.

Рассмотрим этот тезис подробнее. Пусть общее допустимое для проектирования и анализа дерево декомпозиции бизнес-процесса может включать не более $T$ элементов декомпозиции (рассматриваем этот параметр как ресурсное ограничение). Введем еще два параметра: $\lambda$ – среднее число элементов декомпозиции бизнес-функции, и $K$ – глубина дерева декомпозиции. Тогда

$$T = \frac{\lambda^{K+1}-1}{\lambda-1}.$$

Из последнего получаем

$$K = \left\lceil \frac{ln(1+(\lambda-1)T)}{ln(\lambda)} \right\rceil - 1.$$

Как видно из графика этой функции, рассчитанного для $T = 1000$ (Figure 1), с ростом $\lambda$ глубина декомпозиции убывает.

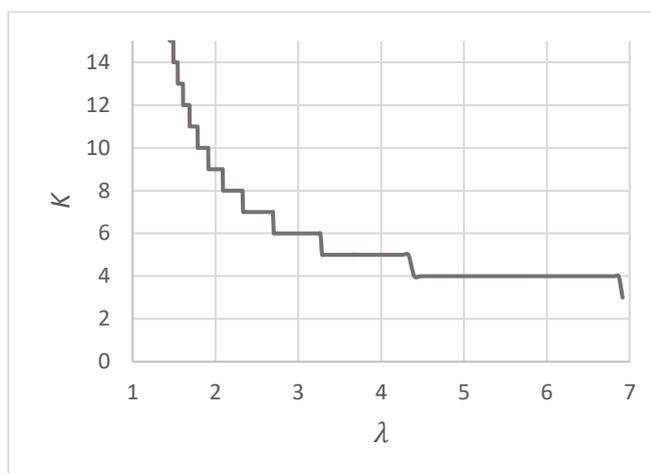

**Рисунок 10. Зависимость глубины декомпозиции от среднего числа элементов декомпозиции**

Однако, можно ли заранее предполагать, какое количество элементов будет в модели? Ведь, по сути, $T$ случайная величина.

А как зависит глубина декомпозиции от математического ожидания количества элементов декомпозиции бизнес-функции?

Мой личный опыт и опыт моих коллег подсказывают, что, как ни старайся, глубина дерева декомпозиции основной массы бизнес-процессов не превышает 4 – 5. Почему?

И еще: почему работает магическое число 7, предложенное Дугласом Россом?

Мы попытаемся найти ответы на эти вопросы.

## 2. Декомпозиция бизнес-процесса

Рассмотрим пример декомпозиции одного из самых тривиальных бизнес-процессов – процесса приготовления яичницы. Для того чтобы описать порядок действий понадобиться построить некоторую умозрительную структуру объектов, в совокупности отвечающей на вопросы «Что делать?» и «Как делать?». Иными словами, надо построить модель бизнес-процесса приготовления яичницы.

Начинаем с объекта, обозначающего бизнес-процесс «Приготовить яичницу». Объекты следующего уровня детализируют действия, необходимые для реализации этого процесса. Будем называть эти объекты бизнес-функциями.

Вообще сам бизнес-процесс имеет ту же природу, что и бизнес-функция. В другом, более общем бизнес-процессе (например, «Приготовить завтрак») этот процесс также назывался бы бизнес-функцией (Figure 2). Мы отделяем бизнес-процесс от бизнес-функций с той целью, чтобы обозначит исходную точку декомпозиции.

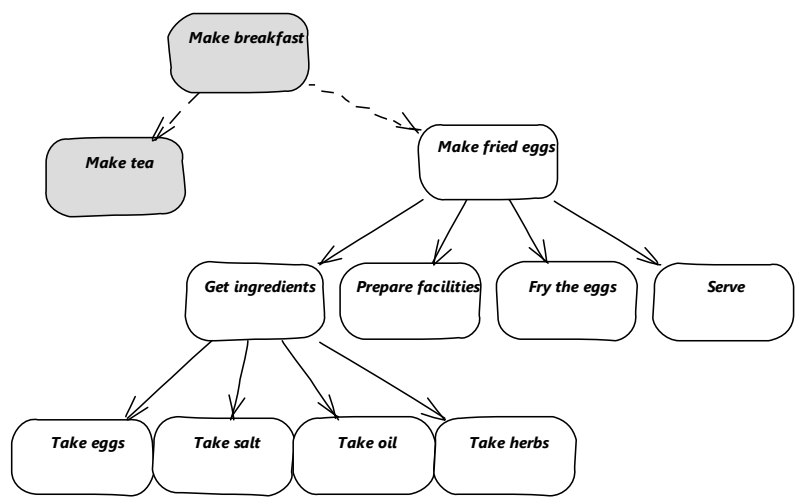

**Рисунок 11. Дерево декомпозиции**

В свою очередь, каждая из представленных бизнес-функций может быть декомпозирована. Например, бизнес-функция «Подобрать ингредиенты» может быть представлена подмножеством таких бизнес-функций: взять яйца, взять соль, взять масло (для жарки), взять зелень (для украшения) и т.д.

В общем случае модель декомпозиции бизнес процесса представляет собой дерево, вершиной которого является наименование бизнес-процесса, а узлами – бизнес-функции.

Процедура декомпозиция бизнес-процесса характеризуется тремя основными свойствами: вариативность, понятность и ограниченность. Рассмотрим их подробнее.

## 2.1. Вариативность декомпозиции

Вариативность определяется как возможность представления декомпозиции любой бизнес-функции различными наборами бизнес-функций. Иными словами, вариативность определяет случайный характер выбора подмножества бизнес-функций, декомпозирующих исходную бизнес-функцию.

Например, в нашем случае можно объединить на первом уровне декомпозиции (Figure 3) бизнес-функции подготовки ингредиентов и приспособлений для изготовления яичницы, назвав новую бизнес-функцию «Подготовиться к приготовлению яичницы», а на следующем, втором уровне, декомпозировать эту функцию на подбор ингредиентов и приспособлений. Выбор бизнес-функций зависит от различных факторов: цели описания бизнес-процесса, понятности описания за счет уменьшения бизнес-функций на каждом уровне декомпозиции, наконец, предпочтения проектировщика.

Важным фактором вариативности является независимость количества элементов декомпозиции любой бизнес-функции от предшествующей декомпозиции бизнес-процесса. Решение о представлении декомпозиции зависит только от выбранной аналитиком детальности описания бизнес-функции и от группировки получившихся элементов декомпозиции.

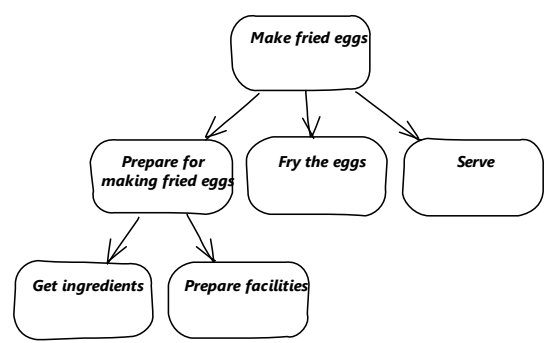

**Рисунок 12. Вариативность декомпозиции**

## 2.2. Понятность декомпозиции

Говоря о декомпозиции бизнес-процесса, автор SADT Дуглас Росс одним из основных требований называл требование понятности представления модели [1]. Он сформулировал эмпирическое правило, по которому декомпозиция любой бизнес-функции не должна включать более 7 элементов.

Декомпозиция не единственный метод представления бизнес-процесса. В предельном случае его можно описать как последовательность операций, приводящих к получению необходимого результата. Однако такое описание для реальных процессов, где количество операций исчисляется сотнями, сложно для понимания и анализа.

Декомпозиция, как метод представления модели бизнес-процесса, создает классификационную схему, группирующую операции в соответствии с системой промежуточных результатов бизнес-процесса. Как отмечал Росс, «хорошая функциональная декомпозиция опирается в первую очередь на хорошую декомпозицию объектов системы».

Таким образом, мерой понятности модели является количество элементов декомпозиции бизнес-функции. Учитывая свойство вариативности декомпозиции, всегда можно добиться, чтобы такое количество было ограничено.

Например, пусть бизнес-функция $A$ декомпозируется на слишком большое (в смысле SADT) количество элементов:

$$A \Longrightarrow (A_1, A_2, \ldots, A_n), 7 \leq n \leq 14$$

Можно ввести дополнительную промежуточную группировку, декомпозировав $A$ на меньшее количество бизнес-функций, которые могут быть декомпозированы далее, то есть

$$A \Longrightarrow (A', A'', A'''),$$
$$A' \Longrightarrow (A_1, \ldots, A_k), k \leq 7,$$
$$A'' \Longrightarrow (A_{k+1}, \ldots, A_m), m - k \leq 7,$$
$$A''' \Longrightarrow (A_{m+1}, \ldots, A_n), n - m \leq 7.$$

## 2.3. Ограниченность декомпозиции

Одним из важных вопросов, решаемых при декомпозиции бизнес-процесса, является определение горизонта декомпозиции, т.е. конечного уровня, на котором следует прекратить декомпозицию бизнес-функций. Максимальный номер конечного уровня декомпозиции называется глубиной декомпозиции бизнес-процесса. Иными словами, глубина декомпозиции – это максимальное расстояние от вершины до листа дерева декомпозиции.

Глубина декомпозиции при построении модели, ее горизонт, определяется либо точкой зрения с которой производится описание, либо предметной областью, в рамках которой описывается бизнес-процесс. Поясним этот тезис.

Описание бизнес-процесса с точки зрения руководителя предприятия должно завершиться до описания производства отдельных комплектующих. Бизнес-функции, детально описывающие это производство, важны для мастера цеха, но не входят в зону непосредственного контроля руководителя предприятия.

Теоретически глубина декомпозиции может быть бесконечно большой: любое действие можно декомпозировать на физиологическом, далее на химическом и физическом уровнях. Однако, совершенно понятно, что в этом случае мы выходим за рамки описания собственно бизнес-процесса и переходим в другую предметную область, которая остается за рамками нашего интереса.

Чем детальнее мы хотим описать бизнес-функцию, тем на более мелкие активности мы должны ее декомпозировать. Например, в нашем случае с функцию «Приготовиться к приготовлению яичницы» можно сразу представить в деталях, указав все необходимые ингредиенты и приспособления конкретно. Таким образом, чем выше детальность, тем больше элементов порождается при декомпозиции бизнес-функции.

Другой пример. При проектировании информационных систем декомпозиция бизнес-процесса завершается на уровне определения производственных операций, которые подлежат автоматизации. С этого момента каждая операция должна быть описана с точки зрения тех информационных сервисов, которые должны быть в ней применены. В этом случае мы переходим в другую предметную область, для которой возникает своя иерархия объектов, описывающих собственно проектируемую информационную систему.

Конечно, общая картина декомпозиции в этом случае представляется единым деревом, но природа элементов этого дерева может быть различной.

### 2.4. Постановка задачи

Резюмируя описание свойств декомпозиции бизнес-процесса, постановка задачи построения модели процесса декомпозиции предполагает следующее.

3) Основанием модели должно служить представление процесса декомпозиции, как стохастического марковского процесса.

4) Модель процесса декомпозиции должна предсказывать следующие вероятностные характеристики:
   – глубина декомпозиции;
   – общее количество элементов декомпозиции модели бизнес-процесса.

### 2.5. Вероятностная модель декомпозиции

С учетом свойства вариативности процесс декомпозиции можно рассматривать как случайный марковский процесс. Рассмотрим формальное определение этого процесса.

Пусть $\mathcal{B} = \{f_i\}$, где $i \in \mathbb{N}$ – множество всех бизнес-функций, относящихся к определенной предметной области, и $B \subset \mathcal{B}$ – некоторое подмножество $\mathcal{B}$. Определим отношение декомпозиции:

$$D_i = D(f_i): f_i \to B_i = (f_{i1}, f_{i2}, \dots, f_{in}), f_i \notin B_i$$
$$\forall i, j \in \mathbb{N}: B_i \cap B_j = \emptyset. \tag{48}$$

Будем говорить, что бизнес-функция $f_g$ принадлежит горизонту декомпозиции, если

$$D(f_g) = \emptyset. \tag{49}$$

Используя данные определения, процесс декомпозиции можно представить следующим образом. Изначально мы имеем один элемент, обозначаемый наименованием декомпозируемого бизнес-процесса:

$$Z(0) = 1. \tag{50}$$

В результате декомпозиции этого элемента в силу вариативности порождается случайное число $X$ элементов декомпозиции первого поколения (уровня):

$$X_1^{(0)} = |D(f_0)| = |B_0 = (f_{01}, f_{02}, \dots, f_{0Z(1)})|. \tag{51}$$

В (4) $X_1^{(0)}$ – количество элементов декомпозиции исходного элемента. Следующие декомпозиции порождаются независимыми декомпозициями каждого элемента предыдущего уровня. Формально декомпозицию можно определить следующим образом:

$$\begin{aligned}
D_0 &= D(0) = (f_0), \\
D_1 &= D(f_0) = (f_1, f_2, \dots, f_n), \\
D_{21} &= D(f_1) = (f_{11}, \dots, f_{1m}); \dots; D_{2n} = D(f_n) = (f_{n1}, \dots, f_{np}), \\
D_{\underbrace{ki\dots j}_{k}} &= D\left(f_{(k-1)\underbrace{i\dots j}_{k-1}}\right) = \left(f_{\underbrace{i\dots j1}_{k}}, \dots, f_{\underbrace{i\dots jp}_{k}}\right).
\end{aligned} \tag{52}$$

Количество элементов декомпозиции элемента $f_{i\dots j}$ k-ого уровня определяется как

$$X_{i\dots j}^{(k)} = |D_{ki\dots j}|. \tag{53}$$

Соответственно, если $f_{i\dots j}$ принадлежит горизонту декомпозиции, то

$$X_{i\dots j}^{(k)} = 0. \tag{54}$$

### 2.5.1. Вероятностное пространство процесса декомпозиции

Элементарной реализацией процесса декомпозиции назовем набор

$$\omega = (f_0, f_{1i}, f_{2ij}, \dots, f_{nij\dots p})/ \tag{55}$$

Вероятностное пространство $(\Omega, \mathcal{F}, P)$ процесса декомпозиции определяется на пространстве событий, представляющем собой множество всех элементарных реализаций

$$\Omega = \{\omega\}. \tag{56}$$

При этом, σ-алгебра $\mathcal{F}$ порождается цилиндрическими подмножествами пространства $\Omega$, а вероятностная мера $P$ задается соотношениями:

$$\begin{aligned}
&P\{X_{i\dots j}^{(k)} = n\} = p_n, \\
&P\left\{\left(X_1^{(0)}, X_i^{(1)}, \dots, X_{i\dots j}^{(k)}\right) = (n_{01}, n_{1i} \dots, n_{ki\dots j})\right\} = p_{n_{01}} p_{n_{1i}} \dots p_{n_{ki\dots j}}.
\end{aligned} \tag{57}$$

### 2.5.2. Декомпозиция как ветвящийся процесс

Учитывая независимость каждой декомпозиции от предыдущих декомпозиций, последовательность случайных величин $Z(0), Z(1), \dots$ представляет собой цепь Маркова.

Шкала времени процесса декомпозиции является дискретной, где последовательные моменты времени соответствуют началу и завершению декомпозиции бизнес-функции. Поэтому можно считать, что время жизни любой декомпозируемой вершины дерева декомпозиции равно 1.

Таким образом, декомпозицию бизнес-процесса можно интерпретировать как ветвящийся процесс Гальтона-Ватсона.

Будем полагать, что декомпозиция описывается надкритическим ветвящимся процессом с ненулевой вероятностью вырождения $\alpha > 0$. Надкритичность процесса означает, что математическое ожидание количества элементов декомпозиции каждой бизнес-функции больше 1

$$\lambda = E[X] > 1. \tag{58}$$

### 2.5.3. Распределение вероятностей для количества элементов декомпозиции

Для определения распределения вероятностей случайной величины $X$, представляющей собой количество элементов декомпозиции бизнес-функции, покажем, что поток событий появления элементов декомпозиции можно интерпретировать как простейший поток, обладающий свойствами стационарности, ординарности и отсутствия последействия [2].

Можно считать, что декомпозиция реализуется на отрезке времени [0,1] и представляет собой случайное небольшое количество элементов. Учитывая особенности процесса декомпозиции, можно утверждать, что элементы появляются на этом отрезке в случайные моменты времени поодиночке. Точнее, вероятность появления одного события на интервале длительностью $\Delta t$ равна $\lambda \Delta t + o(\Delta t)$, где $\lambda$ – интенсивность потока, а вероятность появления на этом интервале более одного события определяется малой величиной $o(\Delta t)$. Это соответствует свойству ординарности потока событий.

Свойство стационарности означает, что вероятность наступления события зависит только от интервала времени $\Delta t$ и не зависит от момента времени $t$. Это верно для процесса декомпозиции, так как элементы декомпозиции, также представляющие собой бизнес-функции, не зависимы друг от друга и, соответственно, каждый элемент может возникнуть в произвольный момент времени на временном отрезке жизни декомпозируемого элемента.

Наконец, независимость элементов декомпозиции позволяет утверждать, что количество событий, возникающих на непересекающихся интервалах времени жизни декомпозируемого элемента также независимы. Этот факт соответствует свойству отсутствия последействия в простейшем потоке событий.

Как показано в [2], вероятность возникновения $n$ событий на временном интервале $\Delta t$ определяется распределением Пуассона, т.е.

$$P\{X(\Delta t) = n\} = ((\lambda \Delta t)^n / n!) \ exp(-\lambda \Delta t). \tag{59}$$

Учитывая, что нас интересует распределение вероятностей для количества элементов декомпозиции бизнес-функции, когда $\Delta t = 1$, то (13) примет вид

$$P\{X = n\} = (\lambda^n / n!) exp(-\lambda), \tag{60}$$

где $\lambda$ – математическое ожидание числа элементов декомпозиции бизнес-функции, соответствующее условию (12).

# 3. *Количественные оценки декомпозиции*

## 3.1. Модель процесса декомпозиции

Рассмотрим следующую модель процесса декомпозиции (см. Figure 4).

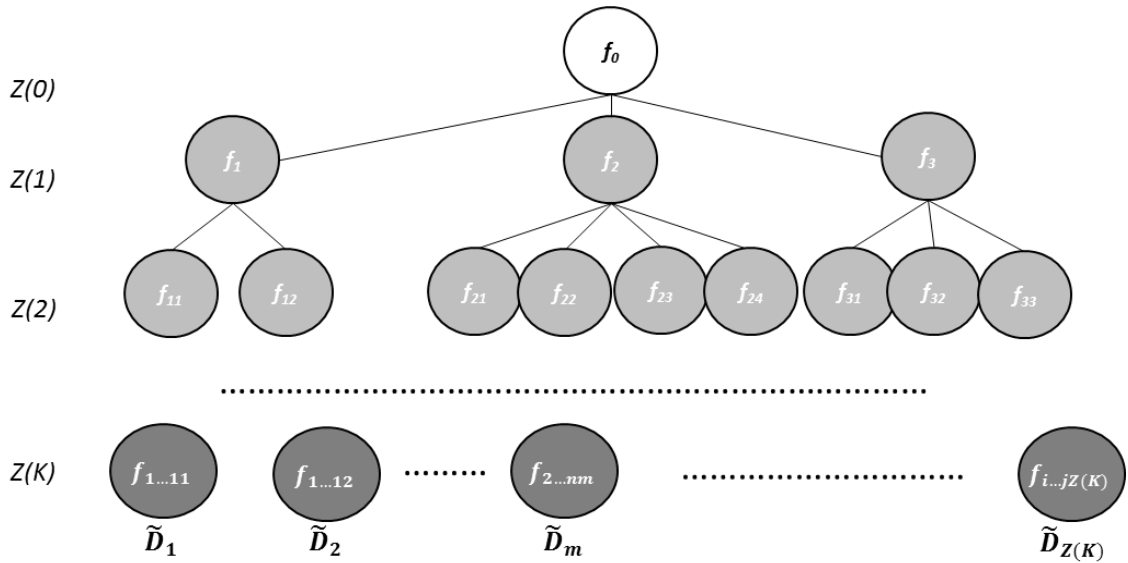

**Рисунок 13. Модель завершения процесса декомпозиции**

Мы будем рассматривать достижение горизонта декомпозиции как вырождение ветвящегося процесса.

Пусть вырождение процесса декомпозиции начинается с некоторого уровня $K$. Количество элементов декомпозиции на этом уровне является случайной величиной $Z(K)$, математическое ожидание которой мы обозначим как $E[Z(K)]$. Понятно, что чем меньше элементов на уровне, тем выше вероятность того, что все эти элементы одновременно не декомпозируются, т.е. на уровне $K$ количество элементов декомпозиции не должно быть настолько большим, чтобы вероятность вырождения не стала пренебрежимо малой.

Рассмотрим каждый элемент этого уровня как вершину отдельного вырождающегося подпроцесса декомпозиции $\widetilde{D}_i$. Ожидаемое количество таких подпроцессов равно $E[Z(K)]$. Обозначим через $\Delta n$ количество элементов декомпозиции, порождаемых в подпроцессе $\widetilde{D}_i$[1].

Для того, чтобы основной процесс декомпозиции завершился, потребуем чтобы далее в этом процессе не ожидалось появление хотя бы одного элемента, т.е.

$$\Delta N = E[Z(K)]\Delta n < 1. \tag{61}$$

В этом случае $K$ определяет максимально возможный горизонт декомпозиции. Декомпозиция бизнес-процесса может завершиться и на более ранних уровнях. Это зависит от разных факторов – от постановки задачи, от уровня понимания процесса проектировщиком и просто в связи с ограниченностью времени проектирования.

---

[1] Далее по тексту все переменные, относящиеся к подпроцессам $\widetilde{D}_i$ будут обозначаться с тильдой.

Для того, чтобы выполнялось условие (15) естественно предположить, что каждый из подпроцессов $\tilde{D}_i$ вырождается уже во втором поколении таким образом, что

$$\Delta n = E[\tilde{Z}(1); \tilde{Z}(2) = 0] = \sum_{i=1}^{\infty} i P\{\tilde{Z}(1) = i; \tilde{Z}(2) = 0\}. \tag{62}$$

## 3.2. Ожидаемая глубина декомпозиции

Ветвящийся процесс описывается производящей функцией

$$f_n(s) = \sum_{i=0}^{\infty} P\{Z(n) = i\} s^i, \ 0 \leq s \leq 1; n \geq 0. \tag{63}$$

Здесь $f_n(s)$ – производящая функция для n-ого поколения ветвящегося процесса, а $P\{Z(n) = i\}$ – вероятность того, в этом поколении ровно i элементов [2]. Таким образом, вероятность вырождения процесса в $n$-ом поколении

$$f_n(0) = P\{Z(n) = 0\}. \tag{64}$$

Производящие функции поколений ветвящегося процесса связаны итеративно

$$f_0(s) = s, \qquad f_{n+1}(s) = f(f_n(s)). \tag{65}$$

В (19) производящая функция f(s) для пуассоновского распределения вероятностей определяется как [3]

$$f(s) = f_1(s) = \sum_{i=0}^{\infty} P\{Z(1) = i\} s^i = \sum_{i=0}^{\infty} \frac{(\lambda s)^i}{i!} e^{-\lambda} = e^{\lambda(s-1)}. \tag{66}$$

С учетом (20) в случае распределения Пуассона производящая функция для n-ого поколения имеет вид

$$f_n(s) = \underbrace{e^{-\lambda} exp\left(\lambda e^{-\lambda} exp\left(\lambda e^{-\lambda} exp\left(\ldots exp\left(\lambda e^{-\lambda} exp(\lambda s)\right)\right)\right)\right)}_{n}. \tag{67}$$

### 3.2.1. Ветвящийся процесс с ожидаемым вырождением

Рассмотрим общую задачу определения математического ожидания числа частиц в $n$-ом поколении ветвящегося процесса с ожидаемым вырождением процесса на уровне $n+1$. Требуется определить для любого $n$

$$E[\tilde{Z}(n); \tilde{Z}(n+1) = 0] = \sum_{i=1}^{\infty} i P\{\tilde{Z}(n) = i; \tilde{Z}(n+1) = 0\} \tag{68}$$

или, учитывая, что $\tilde{Z}(n)$ и $\tilde{Z}(n+1)$ зависимы,

$$\mathrm{E}[\tilde{Z}(n); \tilde{Z}(n+1) = 0] =$$
$$= \sum_{i=1}^{\infty} iP\{\tilde{Z}(n) = i\}P\{\tilde{Z}(n+1) = 0|\tilde{Z}(n) = i\}. \tag{69}$$

Рассмотрим подпроцесс, начинающийся с каждой из частиц *n*-ого поколения. Номер поколения, в котором этот подпроцесс должен выродиться равен 1. А поскольку все $\tilde{Z}(n)$ подпроцессов независимы, вероятность вырождения всех этих подпроцессов одновременно определяется произведением вероятностей вырождения каждого из них. Тогда в правой части выражения (22) условную вероятность можно представить следующим образом:

$$P\{\tilde{Z}(n+1) = 0|\tilde{Z}(n) = i\} = \left(P\{\tilde{Z}(1) = 0|\tilde{Z}(0) = 1\}\right)^i = f_1(0)^i. \tag{70}$$

В (24) использовано выражение для вероятности вырождения ветвящегося процесса (18).

Подставив полученное выражение (24) в (23), получаем

$$\mathrm{E}[\tilde{Z}(n); \tilde{Z}(n+1) = 0] = \sum_{i=1}^{\infty} iP\{\tilde{Z}(n) = i\}f_1(0)^i = f_1(0)f_n'(f_1(0)). \tag{71}$$

### 3.2.2. Максимальный горизонт декомпозиции

Используя (25), определим теперь Δn из выражения (16). Очевидно

$$\Delta \mathrm{n} = \mathrm{E}[\tilde{Z}(1); \tilde{Z}(2) = 0] = f_1(0)f_1'(f_1(0)). \tag{72}$$

Используя выражение для производящей функции распределения Пуассона (21), выражение (26) примет вид

$$\Delta \mathrm{n} = \exp(-\lambda)\left(\lambda\exp(\lambda(\exp(-\lambda) - 1))\right) = \lambda\exp(\lambda(\exp(-\lambda) - 2)). \tag{73}$$

С учетом (27) условие завершения процесса декомпозиции (15) примет вид

$$\Delta N = \lambda^{K+1}\exp(\lambda(\exp(-\lambda) - 2)) < 1. \tag{74}$$

В (28) учтено, что математическое ожидание количества элементов декомпозиции на уровне $K$ равно $\lambda^K$.

Логарифмируя (28) и разрешая неравенство относительно $K$, получаем оценку для максимального поколения вырождения процесса декомпозиции:

$$K < \mathrm{g}(\lambda) = \lambda\left(2 - \exp(-\lambda)\right)/ln\lambda - 1, \tag{75}$$

где $\mathrm{g}(\lambda)$ – функция, определяющая максимальный горизонт.

Таким образом, максимальный горизонт декомпозиции определяется наибольшим целым числом, не превосходящим g(λ):

$$K = \lfloor \lambda (2 - \exp(-\lambda))/ln\lambda - 1 \rfloor. \tag{76}$$

### 3.2.3. Ожидаемый горизонт декомпозиции

Как уже отмечалось, декомпозиция может прекратиться на любом из уровней, предшествующих максимальному. Рассмотрим горизонт декомпозиции как случайную величину $G(K)$, принимающую значения от 1 до K. Задача состоит в оценке возможного закона распределения $P\{G(K)\}$.

Найдем линейное приближение $P\{G(K)\}$, пользуясь следующими соображениями. Будем считать, что

$$P\{G = 0\} = 0 \text{ и } P\{G = K + 1\} = 0. \tag{77}$$

В самом деле, завершение декомпозиции на нулевом уровне не имеет смысла, а вероятность достижения горизонта больше максимального также естественно считать пренебрежимо малой.

Максимально возможным мы будем полагать горизонт близкий к середине интервала $[0, K + 1]$. Как показывает опыт, наиболее вероятным при $\lambda < 7$ является горизонт $G = 3$. Поэтому положим, что мода распределения соответствует значению горизонта

$$K_m = \lceil (K + 1)/2 \rceil = \begin{cases} (K + 2)/2 \text{ для четных } K; \\ (K + 1)/2 \text{ для нечетных } K. \end{cases} \tag{78}$$

Исходя из условий (31) и (32) не трудно получить линейное приближение закона распределения для горизонта декомпозиции:

для четных $K$

$$P\{G = n\} = \begin{cases} 4n/((K + 1)(K + 2)), & 0 \leq n \leq K_m, \\ 4(K + 1 - n)/(K(K + 1)), & K_m + 1 \leq n \leq K + 1, \end{cases}$$

для нечетных $K$ \hfill (79)

$$P\{G = n\} = \begin{cases} 4n/(K + 1)^2, & 0 \leq n \leq K_m; \\ 4(K + 1 - n)/(K + 1)^2, & K_m + 1 \leq n \leq K + 1. \end{cases}$$

С учетом последнего расчет показывает, что ожидаемый горизонт декомпозиции на отрезке $2 \leq \lambda \leq 12$ принимает следующие значения (Figure 5):

$$\overline{K} = \left\lceil \sum_{n=1}^{K+1} nP\{G = n\} \right\rceil = \begin{cases} 3, \text{ при } 2 \leq \lambda \leq 6{,}7; \\ 4, \text{ при } 6{,}7 < \lambda \leq 10{,}7; \\ 5, \text{ при } 10{,}7 < \lambda \leq 12. \end{cases} \tag{80}$$

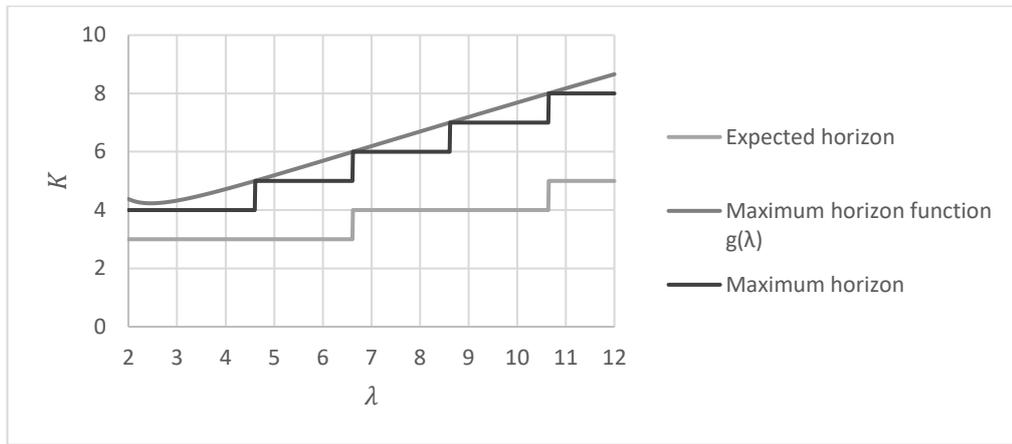

**Рисунок 14. Горизонт декомпозиции**

## 3.3. Оценка общего количества элементов декомпозиции

Для оценки трудоемкости необходимо оценить общее количество элементов процесса декомпозиции. Рассмотрим два сценария такой оценки.

Первый сценарий касается случая полной априорной неопределенности результата декомпозиции, когда заранее невозможно предсказать число уровней модели. Иными словами, сценарий предсказывает общее количество элементов при случайной глубине декомпозиции.

Второй сценарий исходит из наиболее правдоподобной оценки глубины декомпозиции, т.е. когда заранее понятно, какой ожидается горизонт, что бывает в подавляющем числе случаев. Сценарий предсказывает общее количества элементов декомпозиции при ожидаемом горизонте, определяемом соотношением (34).

### 3.3.1. Общее количество элементов при случайном горизонте

Мы полагаем, что глубина декомпозиции является случайной величиной $G$, принимающей значения на интервале $(0; K + 1)$, где $K$ определяется выражением (30). Линейное приближение к функции распределения вероятностей $P\{G = n\}$ задается соотношениями (33).

Обозначим случайную величину общего количества элементов декомпозиции при заданном горизонте $n$ через $T(n)$, а случайную величину общего количества элементов декомпозиции при случайном горизонте - $T(G)$.

Поскольку глубина декомпозиции, как видно из раздела 3.2, слабо зависит от $\lambda$, можно считать, что общее число элементов декомпозиции $T(G)$ и ее горизонт являются независимыми случайными величинами. Иными словами

$$P\{T(n) = j | G = n\} = P\{T(n) = j\} P\{G = n\}. \qquad (81)$$

Тогда вероятность того, что в результате декомпозиции что общее число элементов будет равно $j$ определяется соотношением

$$P\{T(G) = j\} = \sum_{n=1}^{K} P\{G = n\} P\{T(n) = j\}. \qquad (82)$$

Производящая функция для общего числа элементов декомпозиции $T(G)$ соответственно имеет вид

$$C_K(s) = \sum_{n=1}^{K} \sum_{j=1}^{\infty} P\{G = n\} P\{T(n) = j\} s^j = \sum_{n=1}^{K} P\{G = n\} R_n(s), \qquad (83)$$

где $R_n(s)$ – производящая функция общего количества элементов $T(n)$ в процессе с глубиной декомпозиции $n$. Для наших целей конкретный вид функции $R_n(s)$ не имеет значения (рекурсивное соотношение для этой функции приведено в [3]). Наша задача – определить математическое ожидание и дисперсию общего числа частиц процесса декомпозиции $T(G)$, задаваемые соотношениями

$$C'_K(s) = E[T(G)] = \sum_{n=1}^{K} P\{G = n\} R'_n(s)_{|s=1} = \sum_{n=1}^{K} P\{G = n\} E[T(n)], \qquad (84)$$

$$C''_K(s) = E[T^2(G)] - E[T(G)] = $$
$$= \sum_{n=1}^{K} P\{G = n\} R''_n(s)_{|s=1} = \sum_{n=1}^{K} P\{G = n\}(E[T^2(n)] - E[T(n)]). \qquad (85)$$

В (38) математическое ожидание общего числа элементов при глубине декомпозиции $n$ вычисляется как сумма математических ожиданий количества элементов в каждом поколении процесса

$$E[T(n)] = \sum_{i=0}^{n} E[Z(i)] = (\lambda^{n+1} - 1)/(\lambda - 1). \qquad (86)$$

Тогда, используя (38), легко получить выражение для $E[T(G)]$:

$$E[T(G)] = \sum_{n=0}^{K+1} P\{G = n\}((\lambda^{n+1} - 1)/(\lambda - 1)). \qquad (87)$$

Учитывая связь дисперсии с математическим ожиданием

$$D[X] = E[X^2] - E^2[X], \qquad (88)$$

перепишем (39):

$$D[T(G)] + E^2[T(G)] - E[T(G)] = $$
$$= \sum_{n=1}^{K} P\{G = n\}(D[T(n)] + E^2[T(n)] - E[T(n)]). \qquad (89)$$

Откуда, учитывая (38), получаем

$$D[T(G)] = \sum_{n=1}^{K} P\{G = n\}(D[T(n)] + E^2[T(n)]) - E^2[T(G)] \qquad (90)$$

Выражение (44) позволяет рассчитать дисперсию $D[T(G)]$, поскольку все компоненты, необходимые для расчёта определены: $E[T(n)]$ определяется выражением

(47), $E[T(G)]$ – выражением (41), распределение вероятности для горизонта декомпозиции задается соотношениями (33). Дисперсия для общего числа элементов в случае ожидаемого горизонта декомпозиции $D[T(n)]$ определяется выражением

$$D[T(n)] = \lambda^n \frac{\lambda^n - 1}{\lambda - 1} + \frac{2\lambda + 1}{\lambda - 1}\left(\frac{\lambda^{2n} - 1}{\lambda^2 - 1} - \frac{\lambda^n - 1}{\lambda - 1}\right) \tag{91}$$

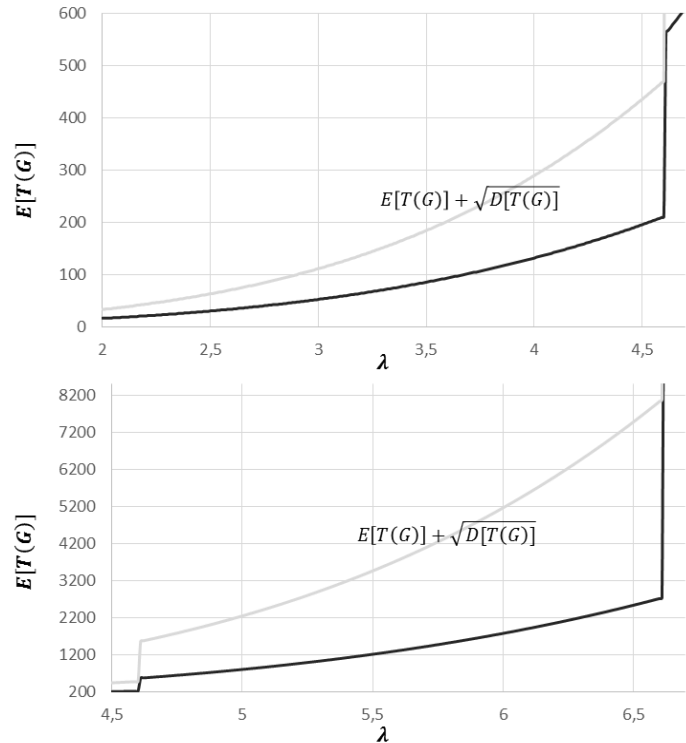

**Рисунок 15. Общее количество элементов декомпозиции при случайном горизонте**

Figure 6 показывает два скачка графика при $\lambda = 4{,}628$ и $\lambda = 6{,}625$. Эти скачки соответствуют изменению ожидаемого горизонта декомпозиции.

### 3.3.2. Общее количество элементов при ожидаемом горизонте

Для *n*-ого поколения ветвящегося процесса (уровня декомпозиции бизнес-процесса) математическое ожидание количества элементов определяется как

$$E[Z(n)] = \lambda^n, n = (0,1,2,\dots) \tag{92}$$

Пусть $T(n)$ – общее количество элементов в процессе, с глубиной декомпозиции $n$. Оценка общего количества элементов декомпозиции для горизонта $\overline{K}$ может быть определена по формуле суммы геометрической прогрессии

$$E[T(\overline{K})] = \sum_{i=0}^{\overline{K}} E[Z(i)] = \frac{\lambda^{\overline{K}+1} - 1}{\lambda - 1}, \tag{93}$$

где $\overline{K}$ определяется выражением (30).

Дисперсия для общего числа элементов в случае ожидаемого горизонта декомпозиции $\overline{K}$ определяется выражением (45).

На графиках (Figure 7) представлена зависимость общего числа элементов декомпозиции от математического ожидания количества потомков единичной декомпозиции на интервалах $2 \leq \lambda \leq 6{,}7$ и $6{,}5 \leq \lambda \leq 10{,}7$.

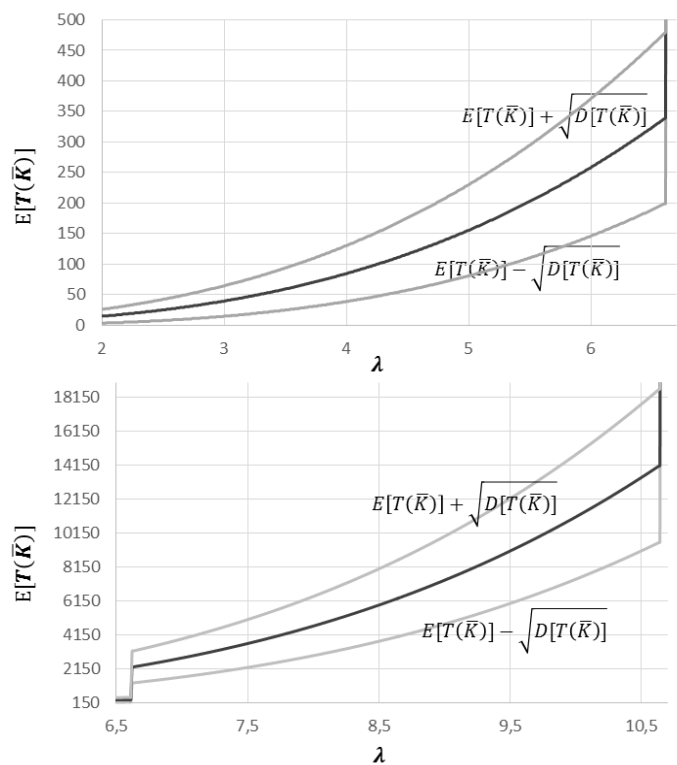

**Рисунок 16. Общее количество элементов декомпозиции при ожидаемом горизонте**

Скачки на графиках соответствует изменению ожидаемого горизонта декомпозиции при $\lambda = 6{,}605$ и $\lambda = 10{,}655$.

## *4. Проверка модели на экспериментальных данных*

В качестве экспериментальных данных использованы декомпозиции бизнес-процессов, выполненных автором в следующих пяти проектах информационных систем:

6) концептуальное проектирование технически сложных объектов;
7) дистанционное обучение детей;
8) аналитическая система авиационной компании;
9) администрирование облачного сервиса;
10) планирование и мониторинг производственных процессов судостроительных предприятий.

Для каждого из проектов формировалась выборка распределения декомпозиций элементов (см. Table 1), на основании которой методами статистической обработки[2] проверялась гипотеза о распределении декомпозиций по закону Пуассона, а также определялось генеральное среднее для единичной декомпозиции.

| Номер проекта | 1 | 2 | 3 | 4 | 5 |
| --- | --- | --- | --- | --- | --- |

---

[2] Статистическая обработка экспериментальных данных выполнялась при помощи интернет-ресурса http://math.semestr.ru/group/poisson-examples.php.

| Элементов в декомп. | | | | | |
|---|---|---|---|---|---|
| **2** | 8 | 6 | 8 | 2 | 4 |
| **3** | 5 | 9 | 10 | 5 | 6 |
| **4** | 7 | 10 | 6 | 3 | 0 |
| **5** | 7 | 13 | 4 | 5 | 3 |
| **6** | 7 | 3 | 2 | 2 | 1 |
| **7** | 3 | 4 | 2 | 2 | |
| **8** | 3 | 3 | | 2 | |
| **9** | 4 | | | 1 | |
| **11** | 2 | | | 1 | |
| **12** | 1 | | | | |
| **13** | 1 | | | | |
| Общее количество элементов декомп. | **264** | **214** | **117** | **118** | **48** |

**Таблица 3. Выборка распределения величин декомпозиций по проектам**

Проверка гипотезы о распределении декомпозиций по закону Пуассона дает положительный результат для всех проектов.

Table 2 и Figure 8 приводят сравнение результатов статистической обработки с модельными расчетами.

| Номер проекта | Экспериментальные данные | | | | | Данные модельного расчета | | |
|---|---|---|---|---|---|---|---|---|
| | Горизонт декомпозиции | Среднее для λ | Нижняя граница дов. инт. для λ | Верхняя граница дов. инт. для λ | Общее количество элементов | Горизонт декомпозиции | Нижняя граница дов. интервала | Верхняя граница дов. инт. интервала |
| 1 | 4 | 5,41 | 4,58 | 6,24 | **264** | 2 - 4 | **0** | 462 |
| 2 | 4 | 4,46 | 3,97 | 4,95 | **214** | 2 - 4 | **0** | 291 |
| 3 | 3 | 3,55 | 3,01 | 4,09 | **117** | 2 - 4 | **0** | 170 |
| 4 | 3 | 5,13 | 4,12 | 6,14 | **118** | 2 - 4 | **0** | 405 |
| 5 | 3 | 3,2 | 2,41 | 3,99 | **48** | 2 - 4 | **0** | 134 |

**Таблица 4. Сравнение результатов статистической обработки с модельными расчетами**

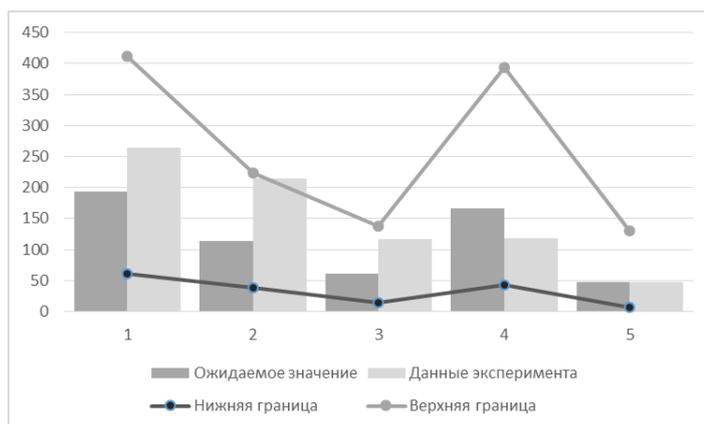

**Рисунок 17. Представление соотношения экспериментальных данных с результатами расчета**

Таким образом, экспериментальные данные для всех проектов соответствуют предсказаниям модели, как по горизонту, так и по общему количеству элементов декомпозиции.

## 5. *Заключение*

Как видим, оценка зависимости глубины декомпозиции, приведенная во введении к данной работе (Figure 1) в корне противоречит реальной (Figure 5): глубина декомпозиции бизнес-процесса растет с увеличением среднего числа элементов декомпозиции бизнес-функции. Попытаемся объяснить этот феномен.

Качество описания бизнес-процесса определяется глубиной и детальностью понимания моделируемого бизнес-процесса аналитиком. Что является мерой такого понимания?

Декомпозиция бизнес-процесса описывается надкритическим ветвящимся процессом Гальтона-Ватсона, который характеризуется ненулевой вероятностью вырождения $\alpha > 0$. Вероятность вырождения – мера понимания бизнес-процесса: чем больше $\alpha$, тем менее детально аналитик описывает бизнес-процесс, тем быстрее завершается декомпозиция. Следуя [4], введем противоположную меру, детальность, $\gamma = 1 - \alpha$. Чем выше детальность, тем лучше аналитик описывает бизнес-процесс.

Вероятность вырождения ветвящегося процесса определяется наименьшим положительным корнем уравнения

$$f(\alpha) = e^{\lambda(\alpha-1)} = \alpha.$$

Откуда математическое ожидание количества элементов декомпозиции бизнес-функции

$$\lambda = -\frac{ln(\alpha)}{1-\alpha} = -\frac{ln(1-\gamma)}{\gamma}.$$

Зависимость математического ожидания $\lambda$ от детальности $\gamma$ представляет Figure 9.

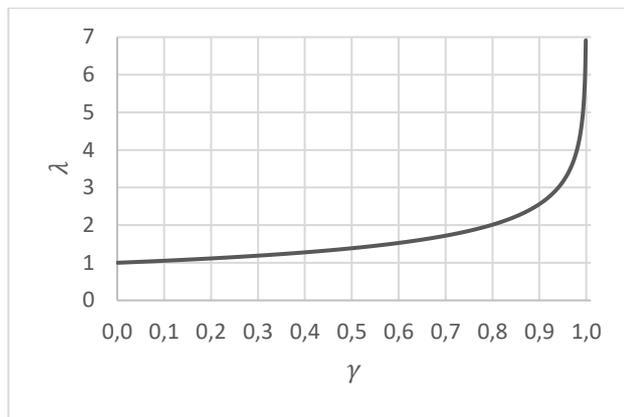

**Рисунок 18. Зависимость математического ожидания количества элементов декомпозиции бизнес-функции от детальности**

Таким образом, рост глубины декомпозиции с ростом среднего числа элементов декомпозиции бизнес-функции объясняется более высокой детальностью понимания бизнес-процесса аналитиком: чем выше детальность, тем выше полнота описания бизнес-процесса.

Figure 9 дает понимание еще одного феномена: магического числа 7, максимального количества элементов декомпозиции бизнес-функции. Как видно из графика, в этой точке детальность понимания бизнес-процесса близка к предельному значению 1 и график выходит на вертикальную траекторию.

## Благодарности

## *6. Список литературы*